\documentclass[preprint2]{emulateapj}
\pdfoutput=1 

\def\arcmin{$^\prime$}
\def\arcsec{$^{\prime\prime}$}
\def\degree{$^{\circ}$}
\def\deg{$^{\circ}$}

\newcommand{\kmsec}{km~s$^{-1}$~}

\usepackage{subfigure}
\shorttitle{The radio continuum structure of Centaurus A at 1.4\,GHz}
\shortauthors{Feain et al.}

\begin{document}

\title{The radio continuum structure of Centaurus A at 1.4\,GHz}


\author{I. J. Feain$^{1}$}\email{ilana.feain@csiro.au}
\author{T. J. Cornwell$^{1}$}
\author{R. D. Ekers$^{1}$}
\author{M. R. Calabretta$^{1}$}
\author{R. P. Norris$^{1}$}
\author{M. Johnston-Hollitt$^{3}$}
\author{J. Ott$^{2}$}
\author{E. Lindley$^{4}$}
\author{B. M. Gaensler$^{4}$}
\author{T. Murphy$^{4,5}$}
\author{E. Middelberg$^{6}$}
\author{S. Jiraskova$^{7}$}
\author{S. O'Sullivan$^{1}$}
\author{N. M. McClure-Griffiths$^{1}$}
\author{J. Bland-Hawthorn$^{4}$}

\affil{1. CSIRO Astronomy and Space Science, PO Box 76, Epping NSW 1710, Australia; ilana.feain@csiro.au}
\affil{2. National Radio Astronomical Observatory, Charlottesville, P.O. Box O, 1003 Lopezville Road, Socorro, NM 87801-0387 , USA  }
\affil{3. School of Chemical and Physical Sciences, Victoria University of Wellington, PO Box 600, Wellington, New Zealand}
\affil{4. Sydney Institute for Astronomy, School of Physics, The University of Sydney, NSW 2006, Australia}
\affil{5. School of Information Technologies, The University of Sydney, NSW 2006, Australia}
\affil{6. Astronomisches Institut der Universit\"at Bochum, Universit\"atsstr. 150, 44801 Bochum, Germany}
\affil{7. Department of Astrophysics/IMAPP, Radboud University Nijmegen, P.O. Box 9010, 6500 GL Nijmegen, The Netherlands}


\begin{abstract}
A 45 deg$^2$ radio continuum imaging campaign of the nearest radio galaxy, Centaurus A, is reported. Using the Australia Telescope Compact Array and the Parkes 64\,m radio telescope at 1.4\,GHz, the spatial resolution of the resultant image is $\sim$600\,pc ($\sim50$\arcsec), resolving the $\gtrsim$500\,kpc giant radio lobes with approximately five times better physical resolution compared to any previous image, and making this the most detailed radio continuum image of any radio galaxy to date. In this paper, we present these new data and discuss briefly some of the most interesting morphological features that we have discovered in the images. The two giant outer lobes are highly structured and considerably distinct. The southern part of the giant northern lobe naturally extends out from the nothern middle lobe with uniformly north-streaming emission. The well known northern loop is resolved into a series of semi regular shells with a spacing of approximately 25\,kpc. The northern part of the giant northern lobe also contains identifiable filaments and partial ring structures. As seen in previous single dish images at lower angular resolution, the giant southern lobe is not physically connected to the core at radio wavelengths. Almost the entirety of the giant southern lobe is resolved into a largely chaotic and mottled structure which appears considerably different (morphologically) to the diffuse regularity of the northern lobe. We report the discovery of a \textit{vertex} and a \textit{vortex} near the western boundary of the southern lobe; two striking, high surface-brightness features that are named based on their morphology and not their dynamics (which are presently unkown). The vortex and vertex are modelled as re-accelerated lobe emission due to shocks from the AGN itself or from the passage of a dwarf elliptical galaxy through lobe. Preliminary polarimetric and spectral index studies support a plasma reacceleration model and could explain the origin of the Faraday rotation structure detected in the southern lobe. In addition, there are a series of low surface brightness wisps detected around the edges of both the giant lobes.  \end{abstract}



\section{Introduction}
\label{intro}

Centaurus A is the radio source associated with the massive elliptical galaxy NGC\,5128. At a distance of $3.8\pm0.1$\,Mpc \citep{harr10acena}, it is by far the closest radio galaxy in the Universe, about five times closer than Virgo A (M87) and over 10 times closer than expected based on the 1.4\,GHz radio luminosity function \citep{mau07}. Since its discovery \citep{bol48,bol49}, Centaurus A has been extensively studied over the entire electromagnetic spectrum, and at a wide range of sensitivities and resolutions. See \citet{isr98} and \citet{gar66} for comprehensive scientific reviews and interesting historical perspectives. A contemporary overview of recent work can also be found in the following series of papers; \citealp{wood10cena,stru10cena,quil10cena,robe10cena,stei10cena,clay10cena,neum10cena,harr10acena,morg10cena,harr10bcena,kach10cena,burt10cena}. \newline

Centaurus A is a Fanaroff-Riley class I \citep{fan74} radio galaxy with a radio luminosity of $L_{\rm 1.4GHz}=2.3\times10^{24}$\,W\,Hz$^{-1}$; \citet{cpc65}. 
Its radio emission is complex and highly structured on all scales; see, for example, radio continuum montages in Figure~1 of \citet{mor99} or Figure~11 of \citet{bur83}. On the smallest scales is a compact (800-1700\,AU, \citealp{gri75,kel97}) radio core associated with the active galactic nucleus (AGN) at the centre of NGC\,5128. A pair of asymmetric nuclear jets are also present, on scales of 1pc (65\,mas) and have been studied in detail using Very Long Baseline Interferometry (VLBI) techniques \citep{pre83,jon96,ting98}. Beyond the nuclear jets are a pair of inner jets that each extend roughly 1.4\,kpc from the nucleus before terminating in lobes that extend a further 5\,kpc from the nucleus \citep{bur83,cla92}. \newline

Emerging from the outer ridge of the northern inner lobe is a {\it middle} jet which extends a further 7\,kpc \citep{mor99} and connects to the northern middle lobe with a scale size of about 14\,kpc. To date no southern radio continuum counterpart to either the northern middle jet or the northern middle lobe has been detected. \newline


Beyond the inner and middle lobes of Centaurus A are a pair of giant outer lobes that extend more than 500\,kpc in projection ($5$\degree$\times\,9$\degree) and account for 73\% of the total radio luminosity \citep{alv00}. The outer lobes have been studied at low angular resolution with single dish and space-borne telescopes, over a wide frequency range; from 4.7\,MHz \citep{ellis66} to 41GHz \citep{isr08}. 
Table 2 in \citet{isr98} gives a good summary of published radio continuum observations of Centaurus A on the various scale sizes described above; see also Tables 1-7 in \citealt{alv00}. \newline

\subsection{The first high-resolution image of Centaurus A}
Up until now, only the central $\sim$1\% of Centaurus A, incorporating the nuclear region, inner and middle jets and lobes, has been imaged at high resolution by using radio interferometric techniques. In this paper we report wide-field (mosaiced) Australia Telescope Compact Array (ATCA) observations of the entirety of Centaurus A at 1.4\,GHz. We briefly describe the observations and image processing techniques, and --- by combining these new data with a reprocessed, archival Parkes 64\,m single dish image at 1.4\,GHz --- we present the first high-resolution ($\sim$50\arcsec) radio continuum images of the giant lobes.\newline  

In this paper, all reported positions are given in equatorial units using the J2000 coordinate system.
\section{Observations}

\subsection{ATCA}
The ATCA was used in mosaic mode over four epochs between 2006 December and 2008 March to observe a total field of view covering 5\degree$\times\,$9\degree\  (in 406 pointings) and centred on $13^{\rm h}25^{\rm m}27.6^{\rm s}$ $-$43\degree01\arcmin09\arcsec. \newline

High dynamic range imaging of complex and extended radio sources ---like Centaurus A--- requires extremely dense sampling of the aperture plane. With no a priori information about the structures in the outer lobes on scales $\lesssim4$', we took great care in our choice of the array configuration. A maximum baseline of 750\,m was chosen as a trade-off between maximising the angular resolution (approximately 50\arcsec\ with Briggs weighting) and maximising the surface brightness sensitivity. Observations were carried out with the four complementary 750\,m array configurations (750A, 750B, 750C, 750D). These configurations were specifically designed for the old\footnote{The advent of the Compact Array Broad Band correlator (http://www.narrabri.atnf.csiro.au/observing/CABB) means that for continuum studies utilising multi-frequency synthesis, the use of multiple array configurations to fill the aperture plane are normally no longer required.} $2\times128$\,MHz correlator to fill the aperture plane in an optimal way. The minimum physical interferometer spacing of the observations was 31\,m, but the minimum effective spacing recovered by observing in mosaic-mode \citep{eke79} was actually about 20\,m, corresponding to a largest angular size of 37\arcmin\ to which our interferometric observations were sensitive.  \newline

\begin{figure*}\centering
\includegraphics[width=15cm]{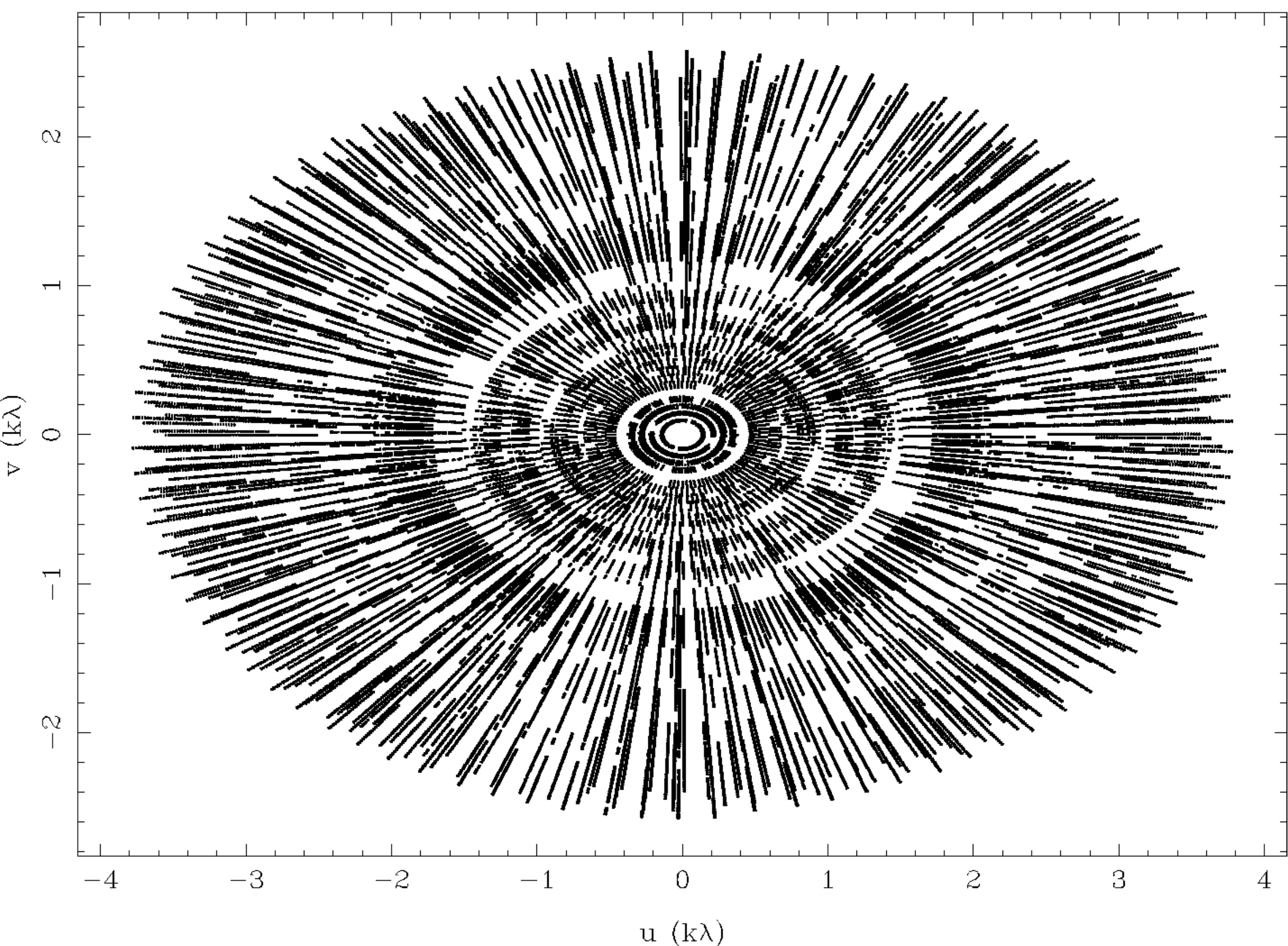}
\caption{\small ATCA observing coverage in the {\textit{uv}}-plane for a typical pointing in the 406-pointing mosaic. The central hole in the {\textit{uv}}-coverage is filled by combining the ATCA observations with data from the Parkes 64\,m telescope.}
\label{image:uvcoverage}
\end{figure*}

Each of the 406 pointings, in each of the four array configurations, received approximately 30-40 `snapshot' cuts over a 12-13 hour period, and we integrated for about 40-45 seconds per cut. A full description of the observations and data calibration have already been given in \citet{fea09}; a summary of the observational parameters is given in Table~\ref{telescopepars} and Figure~\ref{image:uvcoverage} shows the typical \textit{uv}-coverage of the observations. \newline



\begin{deluxetable}{ll}
\normalsize{\normalsize}
\tablecaption{Journal of ATCA observations\label{telescopepars}}

\tablewidth{0pt}
\tablehead{
\colhead{Parameter} & \colhead{Value} \\
}
\startdata
Central Observing Frequencies (MHz) & 1344, 1432\\
Array Configurations & 750 A--D\\
Bandwidth (MHz) & $2\times 128$\\
Channels & $2\times32$ \\
Spectral Resolution (MHz) & 7.08 \\
Synthesised Beam$^a$ & 60\arcsec$\times$40\arcsec, P.A.$=$0\deg\  \\
No. of pointings & 406 \\
Pointing Separation & 16\arcmin\\
Pointing snapshot (sec) & 40\\
Point source sensitivity$^b$ (mJy~beam$^{-1}$)& 0.2-0.3\\
Brightness sensitivity$^b$ (mK)& 25-35\\\enddata
\tablecomments{$^a$Briggs Weighting of visibilities. $^b$Sensitivity reached in outer lobe regions, not near the core where the image is severely dynamic range limited. }
\end{deluxetable}

\begin{deluxetable*}{lcccccc}
\normalsize{\normalsize}
\tablecaption{Measured parameters of some features in the giant lobes of Centaurus A\label{radiofluxes}}

\tablewidth{0pt}
\tablehead{
\colhead{Structure} & \colhead{S$^a$}& \colhead{$\int{S}$$^b$} & \colhead{LAS (R)$^c$} & \colhead{EF$^d$} &\colhead{Thickness (r)$^e$}\\
\colhead{see Fig~\ref{image:real}} & \colhead{mJy~bm$^{-1}$}& \colhead{Jy} & \colhead{arcmin} & \colhead{} &\colhead{arcmin}\\
}
\startdata

Northern Lobe & \\
\hspace{0.5cm}Ring & 39&3 & 20 &1.3&4  \\
\hspace{0.5cm}Shells&25&\nodata&\nodata  &1.3--1.4 &5\\
\hspace{0.5cm}Filament &19& 5&60&1.5&9 \\
\\
\hline
\\
Southern Lobe& \\
\hspace{0.5cm}Vertex&44& 13&30&1.3--1.6&5  \\
\hspace{0.5cm}Vortex & 58&14&45&1.3--1.8&4 \\
\hspace{0.5cm}Wisps & 12&\nodata&\nodata&1.3--1.5&3\\
\tablecomments{$^a$Peak flux density at 1.4\,GHz in units of mJy~beam$^{-1}$. $^b$Integrated flux at 1.4\,GHz in units of Jy. $^c$Largest Angular Size, or R for the purposes of \S\ref{radiophoenix}. $^d$Enhancement Factor (see \S\ref{describe:enhancement}). $^e$Thickness, or filamentary diameter r for the purposes of \S\ref{radiophoenix}.}
\end{deluxetable*}



\subsection{Parkes 64\,m}\label{parkes}
The low spatial frequency (single-dish) image used in this paper is a combination of an unpublished 1.4\,GHz Parkes image (courtesy Norbert Junkes, MPIfR) and a 1.4\,GHz Parkes image created by reprocessing continuum data from the \textsc{hi} Parkes All Sky Survey (HIPASS) survey (Calabretta et al. in prep 2010). Briefly, HIPASS \citep{bar01} was conducted with Parkes between 1997 and 2001, primarily as a blind survey of 21\,cm neutral Hydrogen line emission from galaxies in the local universe. The sky south of declination $+26^\circ$ was scanned five times in declination in 15 zones of $8^\circ$ in width.  The 13-beam Parkes multibeam system was oriented so as to optimize coverage in a single $1^\circ$\,min$^{-1}$ scan, and successive scans at the same declination were stepped so that {\em each} of the $13 \times 14.4$\arcmin\ beams mapped the sky at slightly below the Nyquist rate over five scans.  The dual-polarisation receivers were tuned to 1394.5\,MHz with 64\,MHz bandwidth, and the average system temperature at elevation $55^\circ$ was $21$\,K. \newline

\section{Image Processing}

The following image processing procedure was implemented, after considerable experimentation with various other image processing techniques:

\begin{itemize}
\item Each of the 406 pointings were processed separately using a multi-scale \textsc{clean} deconvolution algorithm \citep{cor08}.
\item A single interferometric image was constructed by linearly combining all 406 individually deconvolved images, weighting by the antenna primary beam and normalising by the sum squared of the primary beam \citep{cor88}. This produced an image of correct flux scale but with noise level increasing at the edge of the sampled area.
\item The interferometric image was combined with the single dish image (\S\ref{parkes}) by feathering in the Fourier plane \citep{Stanimirovic:1999p417}. In this process, the Fourier transforms of the Parkes and ATCA images are added using the Parkes primary beam as a weighting function. 
\end{itemize}

The main problems to be overcome in producing the image were:

\begin{itemize}
\item{\textbf{Brightness of core region:} The peak brightness of the core is about 19\,Jy, whereas the diffuse extended structure of interest ranges from typically $40 - 50$ mJy~beam$^{-1}$ down to $1 - 2$ mJy~beam$^{-1}$. We were unable to reach the theoretical noise level within about 1\deg\ of the core and believe that this is due to residual low level calibration errors ($<1$\%). Self-calibration \citep{Pearson:1984p97} and peeling (e.g. \citealp{mit08}) were unable to bring any improvement. This is not unexpected given the small number of antennas in the ATCA and the complexity of the core region. The effects of the core can be seen even in very distant fields but we were able remedy this by deconvolving two fields jointly - one on the field and one on the core region.}
\item{\textbf{Faint, diffuse structure:} Deconvolution of the faint, diffuse structure in the lobes posed a significant problem. Conventional \textsc{clean}ing with a point source model was unable to extract all the extended emission. Instead, we used a multi-scale \textsc{clean} algorithm \citep{cor08} which models the emission as a collection of different scales. \citet{Rich:2008p7487} have confirmed the efficacy of this approach.}
\item{\textbf{Knowledge of the primary beam of the ATCA antennas:} Joint deconvolution of all 406 pointings could be expected to provide somewhat superior results \citep{cor88}. However, the ATCA primary beam is known with to only about 5\% in the main lobe, which does not allow adequate joint deconvolution.To allow joint deconvolution, we would have to know the primary beam to about 1\%. In addition, we would have to correct for the rotation of the primary beam on the sky. Thus the information on the extended emission comes primarily from the Parkes single dish image.}
\end{itemize}

\section{Results}
\begin{figure*}\centering
\includegraphics[width=11cm]{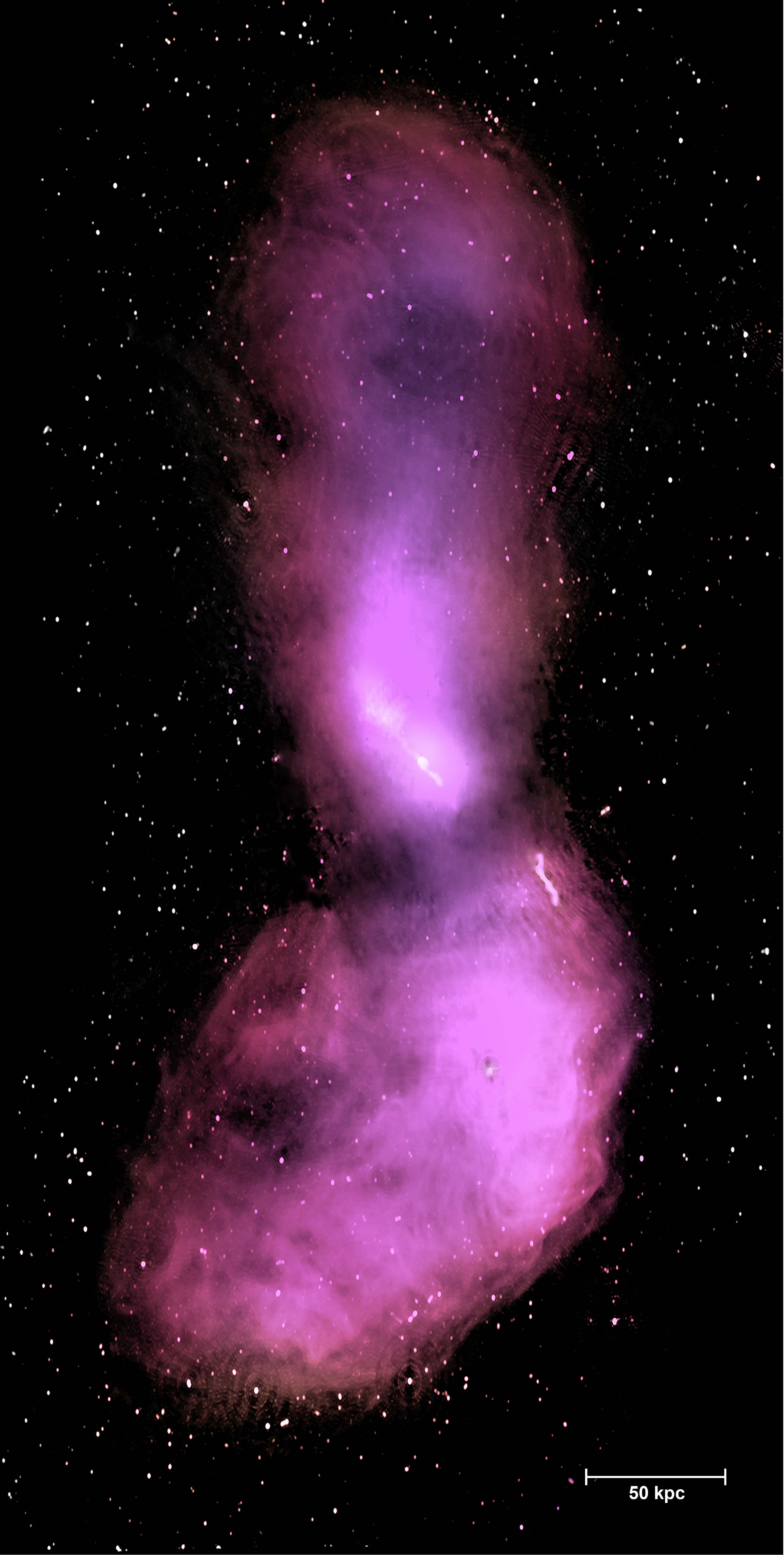}
\caption{\small Enhanced image of Centaurus A at 1.4\,GHz showing the inner lobes, the northern middle lobe (using data from \citealp{mor99}) and the giant outer lobes.This image was created using layering techniques similar to those used by the Hubble Heritage project (http://heritage.stsci.edu), allowing a much improved visualisation of the various scale sizes and structures associated with the inner, middle and outer lobes of Centaurus A, as well as providing an aesthetically suitable image for public outreach. We have taken great care to avoid generating any features in Figure~\ref{image:vision} that are not in our original data. The scientific analysis presented in this paper and elsewhere is, obviously, performed on the original dataset.}
\label{image:vision}
\end{figure*}

\begin{figure*}\centering
\rotatebox{270}{\includegraphics[width=23cm]{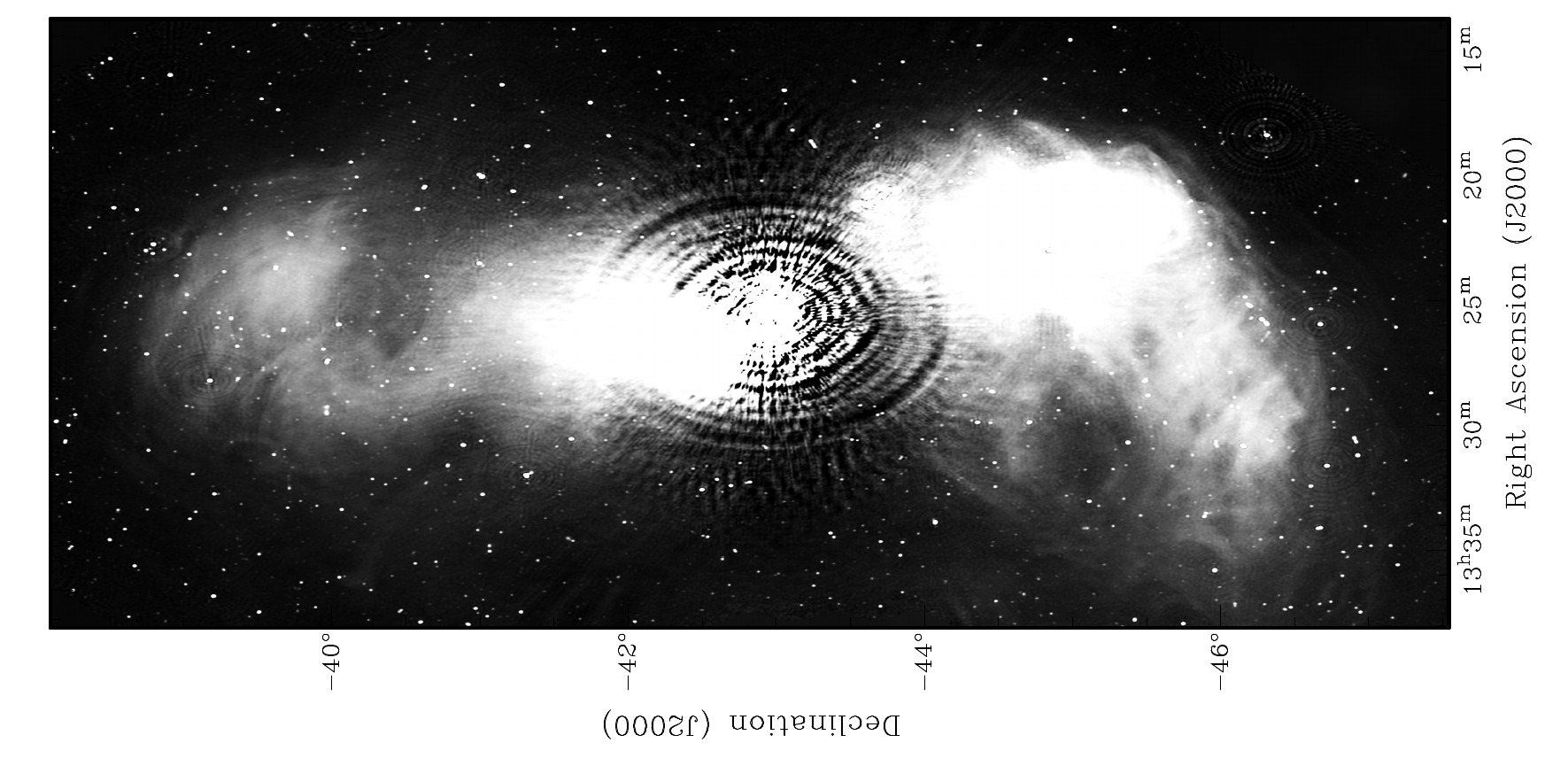}}
\caption{\small 1.4\,GHz radio continuum image of Centaurus A with linear transfer function. The inner jets and lobes and the northern middle lobe are located in the saturated region near the image centre. The angular resolution of this Briggs weighted image is 60\arcsec $\times$ 40\arcsec, P.A.$=$ 0\degree. Residual artefacts in the image are distinguishable from real structure by their obvious radial or tangential origin with respect to the bright nuclear region. }
\label{image:real}
\end{figure*}

For the first time, we have resolved the structures within the giant lobes of Centaurus A. Figures~\ref{image:vision} and \ref{image:real} show different views of the resultant image. Figure~\ref{image:real} is displayed using a linear transfer function that best highlights the giant outer lobes at the expense of saturating the inner regions. The difference in brightness between the two giant lobes themselves means that the best range of intensities to display the (fainter) northern lobe actually saturates resolved features in the southern lobe. \newline 

Figure~\ref{image:vision} was created from Figure~\ref{image:real}, for aesthetic and outreach purposes, using `layers' in gimp\footnote{http://www.gimp.org/} and includes an overlay of the \citet{mor99} 1.4\,GHz ATCA image of the northern middle lobe. The layering technique essentially allows one to display copies of an image with different transfer functions and to overlay these various layers with different prominences in different parts of the image. This technique allows a much-improved visualization of the various brightnesses associated with the inner, middle and outer lobes. Artefacts in the original data (seen in Figure~\ref{image:real}) were removed using a combination of edge filters and interpolation, technically not dissimilar to deconvolution and restoration. We are confident that the scientific validity of the image for the purposes it was intended (public outreach and aesthetics) is retained.

\begin{figure*}
\centering
\includegraphics[width=15cm]{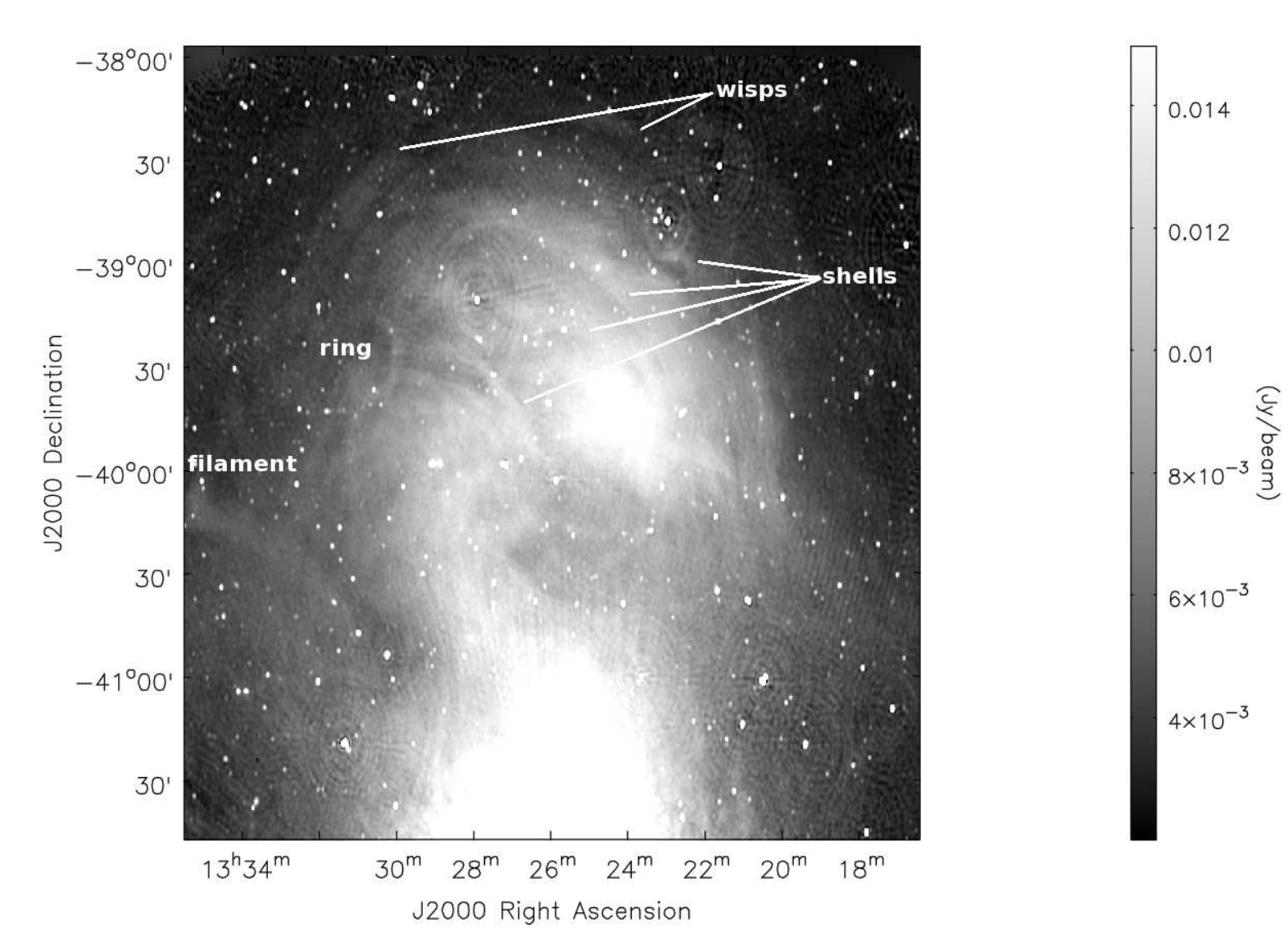}
\caption{\small Close-up view of the northern lobe. Image parameters are the same as those given for Figure~\ref{real}. See \S\ref{describe:northlobe} for details.}
\label{north1}
\end{figure*}

\begin{figure*}
\centering\vspace{2cm}
\includegraphics[width=15cm]{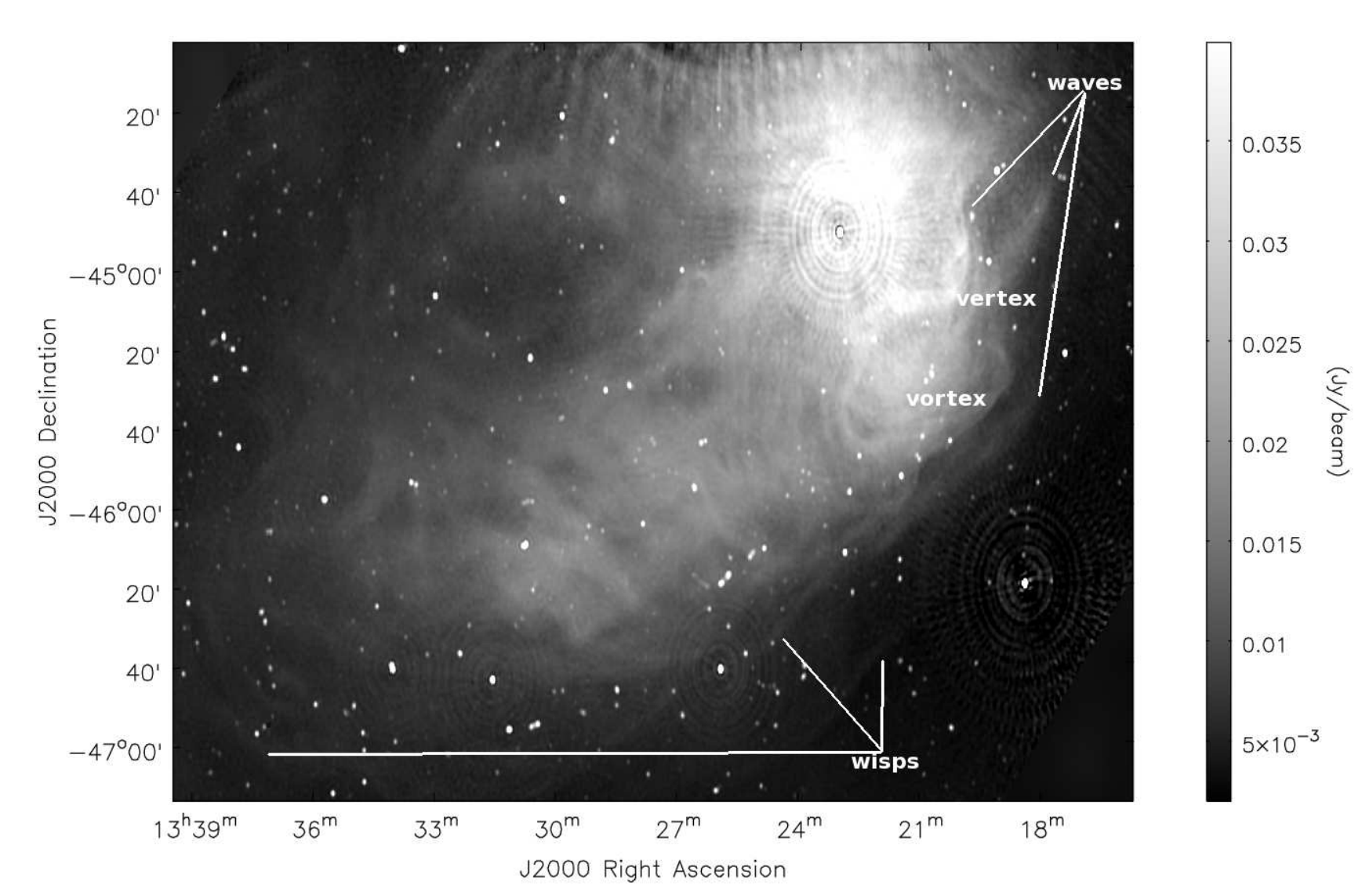}
\caption{\small Close-up view of the southern lobe. Image parameters are the same as those given for Figure~\ref{real}. See \S\ref{describe:southlobe} for details.}
\label{south1}
\end{figure*}

Figures~\ref{north1} and \ref{south1} show close-up views of the northern and southern outer lobes, respectively, and are annotated to identify the structures summarised in Table~\ref{radiofluxes} and further discussed in \S\ref{describe:northlobe} and \ref{describe:southlobe} below. Figure~\ref{image:vertexvortex} is an inset of the vertex/vortex region labelled in Figure~\ref{south1}.\newline

The linearly extended, high surface brightness feature located at $13^{\rm h}21^{\rm m}18.0^{\rm s}$ $-$43\degree41\arcmin21\arcsec\ is the unrelated radio galaxy MRC\,1318$-$434B \citep{sch75,lar81}, associated with NGC\,5090 in the background Centaurus cluster at $z=0.011$. The apparent close alignment between the position angle of the lobes of MRC\,1318$-$434B and that of the inner lobes of Centaurus A is a nice example of cosmic coincidence! MRC\,1318$-$434B is amongst the brightest of the southern radio sources, and probably equivalent in flux density to a 3CRR source \citep{lai83,bur06}. Further analysis of this interesting source is warranted but beyond the scope of this paper. There are also several thousand (mainly background) radio sources discernible in Figure~\ref{image:real}.  \citet{fea09} has published a catalogue of the 1005 compact radio sources in this image down to a flux-density limit of 3\,mJy~beam$^{-1}$, along with a table of Faraday rotation measures and linear polarised intensities for those with high signal-to-noise in linear polarisation. 

\subsection{The northern outer lobe}\label{describe:northlobe}
A close-up view of the northern outer lobe is shown in Figure~\ref{north1}. This lobe appears as a quite natural extension of the northern middle lobe, which itself is spatially connected \citep{mor99} to the northern inner lobe and it eventually to the core. Its overall structure is best described by uniformly north-streaming emission that deviates sharply east into a hook (i.e. the Northern Loop) structure at about  $13^{\rm h}25^{\rm m}00^{\rm s}$ $-$40\degree58\arcmin00\arcsec. The hook itself is resolved into a series of regular shells (\S\ref{describe:shells}) embedded within diffuse, extended emission with noticeable ring (\ref{describe:ring}) and filamentary structures (\ref{describe:filament}). 
\subsubsection{Shells}\label{describe:shells}
There are (at least) three semi-regularly spaced, radially-centric shells that are observed toward the northern extremity of the northern lobe. The innermost shell is the brightest. These shells are $\sim3-6$\, kpc in thickness with an inter-shell separation of $\sim22-28$\,kpc. If the shells are intrinsically physically similar, then variations in shell brightness and thickness and in inter-shell separation would be due to projection effects and could be used in part to start disentangling the three-dimensional structure of the northern lobe. \newline

Possible physical origins for these shells include both intrinsic and environmental causes: the latter reflecting either the underlying density or magnetic field distribution of the intergalactic medium and the former reflecting a possible episodic history of AGN outbursts or of some other periodic outburst of relativistic particles further up the jet, as has been suggested to explain the shells in Hercules A \citep{mas88}. If the shells reflect the local environmental conditions, this implies that there is an enhanced magnetic field up to 1.2 times larger and/or enhanced electron number density up to 1.4 times larger in the shells compared to the intra-shell regions. If, instead, the shells are intrinsic to Centaurus A's emission history, then 3-6\,kpc would correspond to episodic activity from the AGN on timescales of at least 30 kyr (assuming the speed of light and no projection effects). Some splitting of individual shells is also observed, for example at $13^{\rm h}28^{\rm m}$ $-$38\degree45\arcmin, leading to a structure more complex than would be expected from simple periodic outbursts of the AGN. 
\subsubsection{Ring}\label{describe:ring}
Located at approximately $13^{\rm h}30^{\rm m}25^{\rm s}$ $-$39\degree28\arcmin45\arcsec, near the north-eastern edge of the hook, is an elongated partial ring. The location of the ring is indicated on Figure~\ref{image:real}. A possible explanation for the origin of this ring is the creation of a radio `hole' due to the passage of a neighbouring galaxy in the Centaurus group. However, the closest catalogued member of the Centaurus A group is E324$-$24 at $13^{\rm h}27^{\rm m}37.4^{\rm s}$	-41\degree28\arcmin50\arcsec, \citep{kara07} which is located more than 130\,kpc from the ring with a radial velocity of $\sim-30$\kmsec with respect to NGC\,5128. Assuming a typical group velocity of several hundred \kmsec, it would have been at least $\times10^8$ years since E324$-$24 was coincident with the position of the ring.

\subsubsection{Filament}\label{describe:filament}
Protruding out from the eastern boundary of the northern lobe at approximately $13^{\rm h}30^{\rm m}15^{\rm s}$ $-$41\degree06\arcmin10\arcsec, is a diffuse, linear filament which extends approximately 1\deg\ north-east, bending east. The filament, labelled on Figure~\ref{north1}, is most noticeable because it departs so prominently from the overall structure of the northern lobe. The approximate curvature of the filament is similar to that of the shells, which could imply that it, along with the shells, are shaped primarily by the Centaurus group `weather' in a similar way to that seen for some giant radio galaxies \citep{saf09,sub08}



\subsection{The southern outer lobe}\label{describe:southlobe}
A close-up view of the southern outer lobe is shown in Figure~\ref{south1}. Generally, the radio emission is mottled and chaotic, but it includes two embedded or projected high surface-brightness features (the \textit{vertex} and \textit{vortex}; see \S\ref{describe:vertexvortex}) as well as more diffuse, linear radio wisps evident at the south-west and southern extremity of lobe. The central region of the lobe is confused by residual sidelobe structure from the bright background radio source \textsc{pks~1320$-$446}. \newline


The integrated flux of the southern outer lobe is about 30\% larger than the northern outer lobe ($\sim670\,Jy$ versus $\sim500$\,Jy) with no detectable counterpart to the northern middle jet and lobe features. The radio emission from the northern-most tip of the southern lobe decreases sharply into the image noise in what appears to be a physical gap of about 15\arcmin\ (17\,kpc) between the lobe and the core. At a declination of $\delta\approx-44.5$\deg\ low-surface brightness emission from the eastern side of the lobe is probably confused with high-latitude Galactic emission, which is seen clearly in low resolution images extending about 8\deg\ south-east of Centaurus A and bending towards the Galactic plane (\citealp{cpc65,has81}, Calabretta et al. in prep. 2012). \newline

\subsubsection{Vertex and vortex}\label{describe:vertexvortex}
\begin{figure}
\centering
\includegraphics[width=18cm]{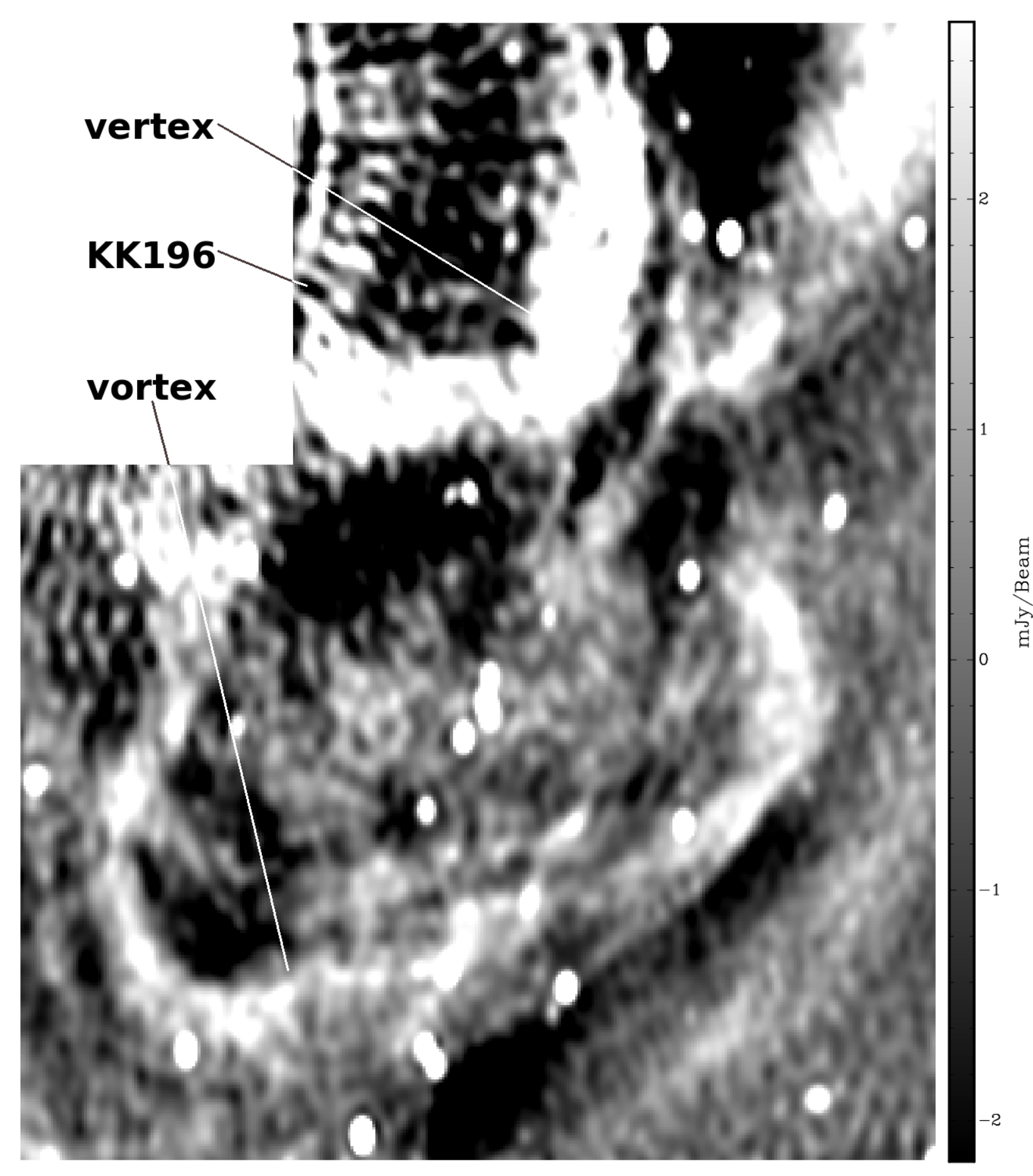}
\caption{\small Close-up view of the vertex/vortex region in the southern giant lobe. See \S\ref{describe:vertexvortex} for details. This is an ATCA-only image which highlights the features more clearly, but at the expense of the extended diffuse lobe emission within which they are embedded. Image parameters are the same as those given for Figure~\ref{real}.}
\label{image:vertexvortex}
\end{figure}

At a position close to $13^{\rm h}20^{\rm m}$ $-$45\degree15\arcmin\ is a bright ridge of radio emission in the shape of a backwards `L', which we have named the \textit{vertex}. Located just below the vertex in the southern lobe is a mushroom-shaped feature, very similar to the feature in the eastern outer lobe of M87 \citep{owe00}, which we have named the \textit{vortex}. A close-up, high angular resolution (ie made using only the ATCA data) view of the vertex/vortex region is shown in Figure~\ref{image:vertexvortex} which better highlights the features at the expense of the diffuse, extended lobe emission that they are embedded within. The vortex structure seen in M87 has been interpreted as the location of termination and backflow of bulk outflow from the AGN. In the case of M87, the bulk outflow is very clearly seen in the radio images. If the vortex/vertex structure in Centaurus A is, similarly, the location of termination and backflow of a bulk outflow from the AGN, then this implies that a bulk outflow (not detected in any radio continuum image thus far) has either recently ceased or is otherwise not radio-continuum bright. \newline 

Interestingly, there is a catalogued member of the Centaurus group that is located within, but close to the eastern edge of, the vertex; its approximate location is marked on Figure~\ref{image:vertexvortex}. The dwarf elliptical galaxy KK196 \citep{kara07,jer00} is at a distance of $3.98\pm0.29$\,Mpc, measured using the tip of the red giant branch, and a radial velocity of $189\pm7$\kmsec relative to NGC\,5128. Thus it seems plausible that KK196 is, or was recently, physically embedded within the southern lobe and may be the catalyst in some way for the production of these structures. This is discussed further in \S\ref{radiophoenix}.


 
\subsubsection{Wisps}
Radially outward from the edges of the northern and southern lobes are a series of increasingly faint wisps, seen most clearly in Figure~\ref{south1}. The physical nature of the wisps is unclear; in the future we will be investigating whether their origin could be similar to the proposed mechanisms for generating wisps in other synchrotron emitting sources like the Crab nebula \citep{gal94,foy07}. The wisps in the northern lobe could be an extension of the shells. Generally, we identify shells as the semi-regularly spaced radially centric structures that make up the northern loop, whereas the wisps are the very low brightness features that propagate outward radially from the edges of the higher surface brightness shells.

\section{Discussion}\label{physical-interpretation}
\subsection{Generation of the vertex and vortex}\label{radiophoenix}

The vertex and vortex in the southern lobe of Centaurus A are similar in size and form to so-called radio phoenixes, a term first coined by \citet{kemp04}, and believed to be the remains of quiescent radio galaxies reaccelerated by their surrounding medium. The prototypical example of a radio phoenix is the highly filamentary object in Abell~85 \citep{slee01}, which displays very similar characteristics to the vertex and vortex over a physical scale of about 100 kpc. Simulations have shown that shockwaves are capable of re-accelerating a fossil electron plasma to produce similar complex, filamentary emission \citep{ens02} produced in this way are expected to have single population spectral indices with strongly polarised filaments. Furthermore at a late stage in the acceleration process the ratio of the global diameter of the radio emission to the filament diameter correlates with the shock strength.\newline

Given the close proximity of the dwarf elliptical galaxy KK\,196 to the vertex and vortex (see \S\ref{describe:vertexvortex} and Figure~\ref{image:vertexvortex}), it is plausible these structures were generated by the passage of KK\,196 through the southern lobe, shocking and re-accelerating a region of its otherwise passively cooling plasma. Alternatively, the features could have been generated from powerful shocks intrinsic to Centaurus A itself, probably originating at its core. The assumption here is that there was, or still is, a southern jet or bulk AGN outflow that is no longer visible and that the vortex structure is the termination point of this jet/outflow. The vertex actually does point straight back to the core. In any case, Centaurus A is (clearly) not a quiescent radio galaxy, but the physics remains the same and we can estimate the required shock strengths to generate these sources from their geometry in the radio image.\newline


A number of authors have considered how to make filamentary radio sources of the order of 100\,kpc to account for the growing number of objects observed (e.g. \citealp{ens02}). Magnetohydrodynamical simulations have shown that there is a strong correlation between the ratio of size of the initial, pre-shocked radio emission (R) and the resultant filamentary structure (r) which arises in latter stages of the shock interaction. Furthermore it has been shown that while the radio plasma will become highly filamentary during a shock passage, the total physical extent of the emisison remains constant with time. Thus, in observing high resolution radio images it is possible to calculate the compression factor which in turn is related to the pressure difference between the pre- and post-shock regions required to generate the emission. Previous efforts to undertake such work have been hampered by lack of sensitivity to both the filamentary and diffuse structure in such sources due to limited \textit{uv}-coverage typical of many radio images which are optimised for either diffuse or filamentary observations, but not generally both. The excellent \textit{uv}-coverage (see Figure~\ref{image:uvcoverage})  of the image presented in this paper makes it one of the few sources where such calculations may be reliably undertaken.\newline

Using the values listed in Table~\ref{radiofluxes}, the ratio R/r is about 11 for the vortex and 6 for the vertex, corresponding plasma to compression factors of about 27 and 8, respectively. Assuming an adiabatic index of 4/3 (for relatavistic gas), this gives pressure differences between the pre- and post-shock regions of 80 and 15, respectively, or if we assume the more likely adiabatic index of 5/3, pressure jumps of 240 and 30, respectively. This can be compared, assuming an adiabatic index of 5/3 (for non-relatavistic gas), with the pressure jump of $\sim$87 and Mach number of $\sim$8.4 associated with the edge of the southern inner lobe \citep{cro09}.
Additionally, \citet{kra03} find a pressure jump of 210 between the shock edge of the southern inner lobe and the more diffuse outer lobe, and similar results have been reported in other systems \citep{min11}. \newline

The vertex and vortex display characteristics both in terms of morphology and energetics to suggest an origin as reawakened lobe plasma either by powerful shocks originating from the Centaurus A core itself, or by the passage of a dwarf elliptical galaxy passing through the southern lobe. Preliminary investigations of the these features appear to support the radio phoenix hypothesis in terms of their radio spectral indices and in fractional polarisation. In future work, we will present a detailed analysis of these features to explore fully their emission mechanism.



\subsection{Shocks and/or instabilities in the southern lobe?}
\begin{figure}
\centering
\rotatebox{270}{\includegraphics[width=10cm]{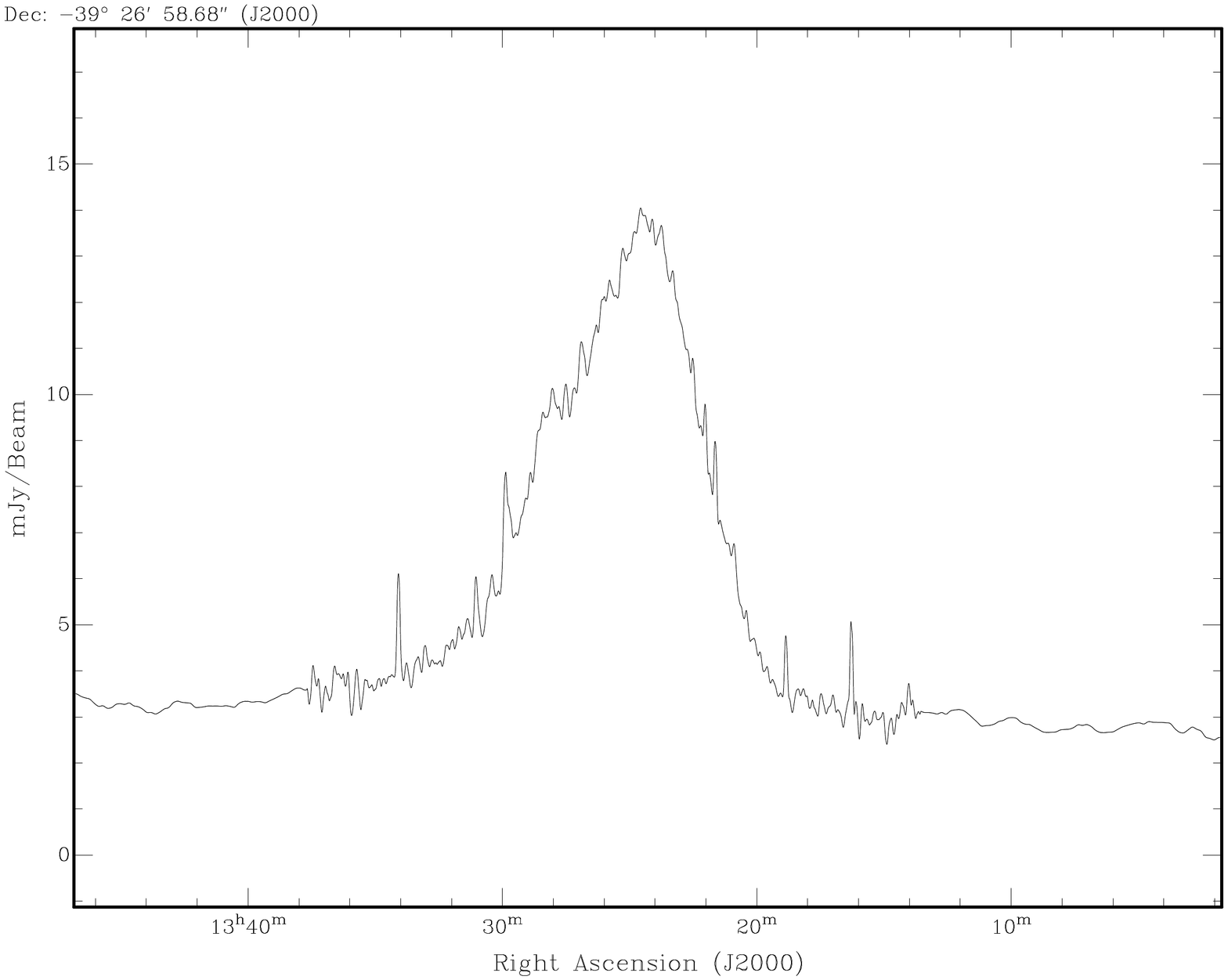}}
\rotatebox{270}{\includegraphics[width=10cm]{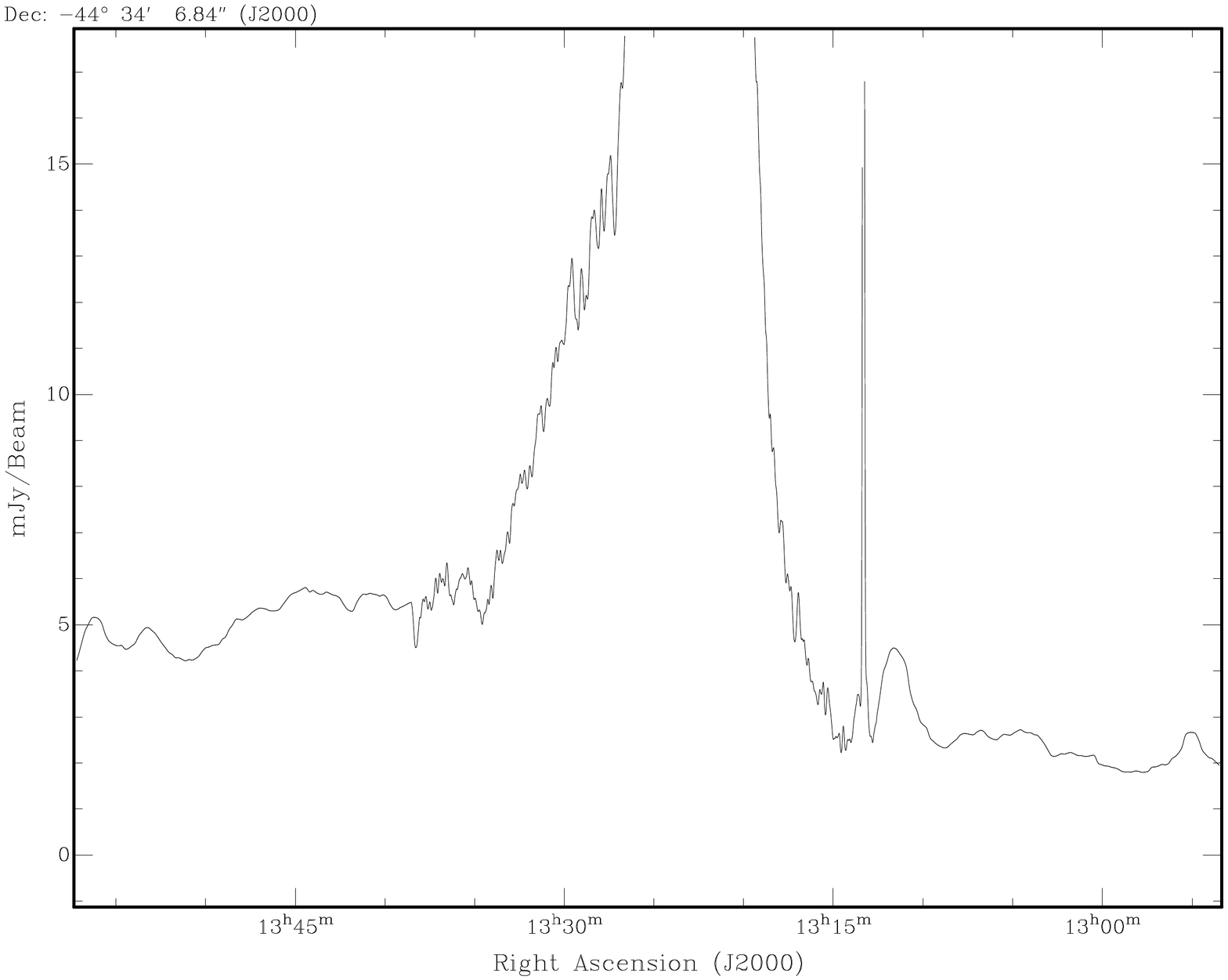}}
\caption{\small Slices in right ascension through the northern (top panel) and southern (bottom panel) lobes to illustrate the existance of a sharp edge on the western side of the southern lobe. The declination at which the slices were taken are given in the top left corner of each panel.}
\label{image:Xslices}
\end{figure}

Figure~\ref{image:Xslices} shows slices in right ascension through the northern and southern lobes. The south-western boundary of the southern outer lobe (the right hand side of the bottom panel in Figure~\ref{image:Xslices}) displays a fairly sharply defined edge, in comparison both to the eastern edge of the southern lobe and to the boundary of the northern lobe. This edge has surface wave-like structures similar to those observed in Cygnus A \citep{bick90}. In the case of Cygnus A, the surface waves are interpreted as Kelvin-Helmholtz instabilities on the lobe-medium boundary, giving rise to a thin skin where turbulent mixing between the lobe and intralobe media occurs and which in turn generates turbulent Faraday rotation along the boundary.\newline 

\citet{fea09} reported the detection of a turbulent Faraday rotation measure (RM) signal, with rms $\sigma_{\textrm{RM}} =17$\,rad~m$^{-2}$ and scale size $0.3$\degree, associated with the southern giant lobe, but no detectable signal from the northern outer lobe. We could not verify whether the signal arose from turbulent structure throughout the lobe or in a thin skin surrounding the lobe boundary; although the latter was favoured. The wave-like features detected along the south-western boundary of the southern lobe, together with the detection of a turbulent RM signal in the lobe, could be due to Kelvin-Helmholtz instabilities. However, given the proximity of the vertex and vortex to the western edge, it is equally plausible that the turbulent RM signal arises mainly from the shocks that generated these features, as discussed in \S\ref{radiophoenix} above. Alternatively, if the vertex/vortex system is simply part of the wave features as described, then it may be possible that this region of the southern lobe is dominated by a backflow system of turbulent mixing. 
\subsection{Common enhancement factors in the lobes of radio galaxies}\label{describe:enhancement}

For each of the features discussed in this paper, and given in Table~\ref{radiofluxes}, we have measured their enhancement factor as the ratio of the typical brightness of the feature to that of the lobe emission the feature is embedded within (or projected onto). In all cases, the enhancement factor, which is given in Column 5 of Table~\ref{radiofluxes}, is approximately 1.3--1.5. This is perhaps surprising given the range of structures, sizes and the physical separation of the features. We note that the enhancement factors measured for similar structures in other radio galaxies are also uniform across the extent of the source; in Table~\ref{table:enhancment} we have tabulated the range of values for a number of well known radio galaxies. Possible explanations for the uniformity of the enhancement factor range from the possiblity of a single underlying physical process generating (and confining) the structures to a projection effect of relatively small structures embedded within a much larger three-dimensional lobe. These filament enhancement factors are smaller by a factor of a few compared to those of knotty structures within a jet boundary, and are, unlike the knots in the jet, not expected be indicative of strong shocks. Parameters such as jet power, magnetic field strengths and matter densities, and in some cases the density of the medium surrounding the lobes may result in slight differences of the enhancement factor between different sources. A better knowledge of relative spectral ages and particle density or magnetic field strengths, together with a better understanding of the three-dimensional structure of the lobes is required before we can interpret the uniformity of the enhancement factor in Centaurus A any further.

\begin{deluxetable*}{lcccl}
\normalsize{\normalsize}
\tablecaption{Estimates of the range of enhancement factors (EF) measured in other radio galaxies\label{table:enhancement}}

\tablewidth{0pt}
\tablehead{
\colhead{Radio Galaxy} & \colhead{EF$^a$} & \colhead{Type$^b$} & \colhead{$\lambda^c$}  & \colhead{Reference} \\
}
\startdata
Centaurus A & 1.3--1.5 & FR\,I & 20\,cm & this paper\\
Fornax A & 1.3--1.5 & FR\,I & 20\,cm & \citealp{fom89}; this paper\\
Virgo A & $\sim$2 & FR\,I & 90\,cm & \citealt{owe00}\\
Cygnus A & 1.5--1.9 & FR\,II & 6\,cm & \citealp{car89,car91}; this paper\\
Hercules A & $\sim2.5$ & FR\,I/II & 20\,cm & \citealt{sax02}\\
B2147+816 & $\sim2$ & FR\,II & 20\,cm & \citealt{kro04}\\
Pictor A & 1.5 & FR\,II & 20\,cm & \citealt{per97}\\
\tablecomments{$^a$Enhancement Factor. $^b$Radio galaxy morphological type in terms of its Fanaroff-Riley classification. $^c$Wavelength of the image used to measure the enhancement factor. }
\end{deluxetable*}

\section{Final Remarks}
After more than 60 years of study, Centaurus A continues to provide us with new insights into almost all areas of astrophysics. It is seen right across the electromagnetic spectrum and may well be the first identified discrete extragalactic source of cosmic rays (e.g. \citealp{pao07,gor08,far08}). It has emission from stars, neutral, molecular and ionised gas, relativistic plasma and a central supermassive black hole with an accretion disk and radio and X-ray jets triggering star formation far beyond the nucleus. This is probably not because Centaurus A is peculiar but because it is so close (10 times closer than it ought to be) and, therefore, it can be studied in unprecedented detail. So, while new technology is allowing us to undertake large area and all-sky surveys of millions of radio sources at many wavebands, it remains both an essential and complementary approach to continue detailed investigations of individual radio galaxies with extremely good sensitivity and high resolution, as a means of fully understanding AGN feedback, and of the evolution of low power radio galaxies in general. What exactly can Centaurus A teach us about the life cycle of low power radio galaxies? Do the time-averaged properties of the best numerical models produce the overall properties of the low power radio galaxy population?\newline

It has become clear in recent years that AGN activity, and in particular so-called `radio-mode' feedback, has important consequences for massive galaxy formation and evolution, but the precise nature and detailed physics underlying these consequences are unclear.  Generally, there are at least two distinct modes of feedback and questions remain as to whether they compete or dominate at different stages of a galaxy's life cycle. Negative feedback, in which the radio jets/lobes heat infalling gas and halt star formation in the most massive galaxies \citep{raw04,cro06,booth09,catt09} is one mode. Positive feedback, in which the jets/lobes shock the infalling-gas, triggering subsequent star formation \citep{wvb85,rees89,dey97,croft06} and perhaps a population of jet-induced galaxies and quasars in the early universe \citep{kla04,elbaz09}, is the other mode. \newline

In many ways, Centaurus A can be considered typical of the dominant (low-luminosity) population of radio galaxies in the universe and in this way its AGN feedback history may be indicative of global radio-mode feedback. It is well established that on scales of the northern middle lobe and smaller, positive feedback is responsible for triggering a large, but not dominant, population of massive stars \citep{gra98,mou00,bla75,mou00,sch94,char00,oos05}. But, given that these inner regions account for less than 27\% of the total energy at radio wavelengths and amount to less than 1\% of the total physical size, we really don't yet understand the overall AGN feedback history of Centaurus A. \newline 

So, 60 years on, we still have a lot to learn from Centaurus A before we have a detailed physical model, from inception to death, of this source and, ultimately, of the low power radio galaxy population as a whole. \newline


\subsection{Summary of this work}
In this paper, we have presented a 45 deg$^2$ radio continuum image of the entire structure of the nearest radio galaxy, Centaurus A. At 1.4\,GHz, the spatial resolution of this image is $\sim$600\,pc, and we have clearly resolved the $\gtrsim$500\,kpc giant radio lobes with approximately five times better physical resolution compared to any previous image of Centaurus A. We have explained the observations and image processing techniques and briefly described the challenges associated with such wide-field, high-dynamic range, imaging. The two giant outer lobes are highly structured and considerably distinct and we have discovered and documented several structures in the northern (shells, filaments and a partial ring) and southern (vertex, vortex and boundary wisps) lobes. The southern lobe remains physically dissociated with the core of Centaurus A, with no discernible counterpart to either the northern middle jet or lobe detected. Yet, the southern lobe is by far the more physically interesting lobe with clear evidence for shocks, turbulence and interactions with the group medium. Specifically, we have idenitified two high surface brightness features that we model as radio phoenixes and whose origin we speculate as due to the powerful shocks created either by the passage of a dwarf elliptical through the southern lobe of Centaurus A, or by the intrinsic energetics of the AGN itself. \newline 

The image presented in this work is publically available through the NASA/IPAC Extragalactic Database (NED) at http://nedwww.ipac.caltech.edu/.\newline

\acknowledgements{The authors wish to thank Katherine Newton-McGee, Minnie Mao and Jay Ekers for helping with the ATCA observations. The Australia Telescope Compact Array and Parkes dish are funded by the Commonwealth of Australia for operation as a National Facility managed by CSIRO. This research has made use of the NASA/IPAC Extragalactic Database (NED) and the NASA Astrophysics Data System (ADS). B.M.G. acknowledges the support of the Australian Research Council through grant FF0561298.}
\bibliography{mnemonic,mnemonic-simple,biblio}
\newpage

\end{document}